\centerline{Thermodynamic Limits, non-Commutative Probability, 
 and Quantum Entanglement}

\centerline{J.~F. Johnson}

Mathematics Department, Villanova University

We construct a rigourous model of quantum measurement.  A 
two-state model of a negative temperature amplifier, such 
as a laser, is taken to a classical thermodynamic limit.  
In the limit, it becomes a classical measurement apparatus 
obeying the stochastic axioms of quantum mechanics.  Thus 
we derive the probabilities from a deterministic Schroedinger's 
equation by procedures analogous to those of classical statistical
 mechanics.  This requires making precise the notion of 
`macroscopic.' 

Macroscopic entanglement is a fact of life.  But the topic of Quantum Measurement has 
to explain in what sense decoherence between pointer positions of a measurement apparatus 
can happen, and propose a mechanism to produce it.  In this paper we study 
a simple two-state model of an amplifying apparatus being used as a particle 
detector.  
Schroedinger's equation's being exactly valid over laboratory dimensions, including the 
amplifying apparatus,
 the states of the particle and the apparatus are entangled, 
after the measurement process is complete.
  We take a closed system approach, so we neglect 
the effect of the environment.  If the apparatus is large enough, it is well 
approximated by a new kind of thermodynamic limit, which we introduce, and the 
Ming Effect shows that superpositions of pointer positions are practically negligible.
No special mechanism is required.

\centerline{Introduction}

  Although Quantum Mechanics, in the form discovered by Heisenberg,  Schroedinger, and Dirac in 1925, is the most successful  physics theory in history, explaining almost every phenomenon within laboratory dimensions,  especially including all of chemistry and biology, there has always been some  dissatisfaction with it.  Nobel prize-winners such as Einstein, de Broglie,  Planck, and Schroedinger himself have all ended up dissenting from the usual  interpretation of Quantum Mechanics.  At issue is not the formal, mathematical  methods of calculating experimentally verifiable quantities, but rather what  philosophical interpretation or mental picture of the wave-particle duality  can be considered true, or even sensible?      Wigner [15] analysed this as the problem of Quantum Measurement.  As long as we do not measure anything,  all Nature is governed by a deterministic wave equation, Schroedinger's equation, and what the wave will do in the future is completely determined  by what it did in the past.  All calculations of experimentally verifiable  quantities involve calculations involving this wave and using the wave equation. But whenever we measure a system, its wave makes some kind of jump which  is a violation of the wave equation, is random and unpredictable, and our  measurement always seems to be the detection of a particle (the quantum of the  wave's field). 
This is no logical contradiction: we never have  trouble calculating predictions and verifying them, although the predictions of  measurements are always merely predictions of the probabilities that the wave  will make the different possible jumps.      
Different views on the subject  make no practical difference to actual experimental design.  
  So, for all practical purposes so far, this puzzle is irrelevant.  On the  other hand, advances in mesoscopic engineering and quantum computation may  upset this quiescent state of affairs. Nevertheless, as  the famous particle physicist, John S.\ Bell remarked more than once,  ``as long as the [jump] is an essential component, and as long as we do not  know exactly when and how it  takes over from the Schroedinger equation, we do not have an exact and unambiguous formulation of our most fundamental theory.''[1]

 \centerline{The precise problem}

The main difference between Quantum Mechanics and Classical Mechanics, ever  since the time of de Broglie, has been that a particle is described  by a wave and a wave equation.  So is a two-particle system.  In fact, any  system, no matter how large, is, if it is a closed system, described by  a wave function and obeys a wave equation, Schroedinger's equation. The next most important difference, and it only was realised a few years later,  by Dirac, Heisenberg, and Born, is that if a closed system is going to be measured, then it is not closed anymore!  It is being impacted by the  measuring apparatus in an unavoidable, essential way.  Therefore, a jump  may occur, and the laws of Quantum Mechanics for measurement processes no  longer is a wave equation, but a way of calculating the probabilities of the  different jumps and the different results of the measurement, as registered  by different pointer positions on the macroscopic measuring apparatus for example.  Although these new rules use the wave function, they violate  Schroedinger's equation, and are needed in order to agree with experiment.

This is new, novel and counter-intuitive, but it is not illogical  or inconsistent.  Most practicing physicists especially in the thirties  could tell when they were measuring a microscopic system and when they  were not, so they ``just knew'' when to use Schroedinger's equation and  when to use the jump rules, which were, in those days, referred to as  `the reduction of the wave packet.'  The convenient consensus grew up, that actually, Schroedinger's equation  only applied to microscopic systems that were so small that they could not be  measured except by amplifying devices.  And that measuring devices such as  amplifying devices---this includes photographic plates, photo-amplfiiers,  lasers, cloud chambers, bubble chambers, Geiger counters, just about everything, ---{\it had } to be macroscopic and so large that Quantum Mechanics was not really  applicable, one could use classical Newtonian Mechanics for them.   Of course this was just a dodge to avoid dealing with an intractable problem. Recent advances in technology have eroded this consensus: macroscopic quantum  tunnelling has been exhibited in the laboratory, mesoscopic quantum chaos has  been detected, and it is no longer believed that one can draw a line between  the microscopic realm and the realm of measurement apparati.  They must be  governed by the same laws of Nature.

Wigner's fundamental article [15] draws out the logic of this situation quite clearly.  According to Quantum Mechanics, there are two seemingly incompatible  ways to analyse the following physical process:  Suppose we have a particle being detected by a Geiger counter or any measurement apparatus that detects  particles.  Suppose further that the counter is tuned so that it will detect  the particle if it is vertically polarised, but not detect it if it is  horizontally polarised.  Now the laws of Quantum Mechanics tell us that the wave function of the incident particle can be either purely vertically polarised, or purely horizontally polarised, or some superposition ({it i.e.,} linear combination) of the two.   On the one hand, this physical process is a measurement.  The combined system of incident particle and measurement apparatus  is a measurement, and therefore the laws of Quantum Mechanics about  measurements say that the particle will have some probability of being  detected and jumping into a state of pure vertical polarisation, and will  have the complementary probability of being absorbed and {\it not } detected.  The probability depends on how much of the pure vertical state  is present in the superposition.  

On the other hand, this combined system of particle plus apparatus is a physical, closed system, governed by the laws of Nature, which are Quantum  Mechanics, which means Schroedinger's equation.  There must be {\it one } wave  function which describes the {\it joint } behaviour of the particle {\it cum } apparatus. This total wave function evolves according to the appropriate wave equation  of the total combined system, and does not jump, and there are no probabilities: it evolves completely deterministically.

Wigner's own solution to this philosophical puzzle has not been widely accepted.   It is that Schroedinger's equation is true as long as consciousness does not  {\it look } at the result of the measurement, but when consciousness is involved,  some non-linear equation modifies the behaviour of the total wave, with  unpredictable results.  He could not propose an actual equation, so this solution  need not detain us.

Now, there is no way to compare a wave function of a large object, such as  an apparatus, with the sense perceptions which we have of that large object,  for example, our conscious visual observation of its pointer position.  The  wave equation allows for the object to have one possible state, say the pointer  position being ``detected!'' and another, quite different, state, say,  the pointer points to ``not detected'' but it also allows for the apparatus to be in a superposition of  two states, e.g., 50-50 one and the other.  No one knows what such a superposition  of states would look like, and current opinion is divided on whether they  can exist, or whether they do exist but decay rapidly into definite `either one or the other but not superposition' states when observed or tugged at or jiggled or something.  But in superconductivity phenomena, such states have  been detected experimentally (but, alas, invisibly, so we still don't know  what they `look' like).

Since there are no rules about what such superpositions look like, or mean macroscopically, there is no logical contradiction between the two different ways to analyse the same physical measurement process.  There is no way to compare the view which cuts off the apparatus from the incident particle, describing one by a wave function but the other classically, with definite pointer positions at all times and never quantum superpositions---with the view which does not cut  Nature anywhere, but studies the joint wave function in its continuous  deterministic evolution.  Since there is no way to confront the two ways  with each other, there is no contradiction.  But since there is no way to compare them either, it is a puzzle.

It seems that there ought to be a way to compare the macroscopic description  of some body, some object, such as a measurement apparatus or our sense-perceptions of it and of its pointer positions, with the microscopic description of the same object, which consists simply  and solely of its wave function.  But there hasn't been, yet.  There has not been a precise definition of macroscopic, and  the rough-and-ready thirties style definition is unconvincing.   Nor has there been a precise definition, in terms of the wave equation, of  measurement process.  In a recent paper,[11] I have proposed a solution to these  two questions which solves the problem of quantum measurement, or at least  that part of it which work in the last decade has concentrated on (the jumps have usually been regarded as [9] a merely subjective selection from the mixture, few now believe that they really happen).

Macroscopic entanglement is a fact of life ({\it pace } W. Zurek [16]).  Therefore, we take 
Schroedinger's equation to be exactly valid over laboratory dimensions, including the 
amplifying apparatus.

Then the problem becomes how to explain mathematically how the wave of the measured system, when interacting with the wave of the amplifying apparatus, produces intermittent particle events such as the click of a Geiger counter.  The production of a discontinuous transition from continuous inputs is often the result of an interchange of limits phenomenon, and that is what we have here.   

\centerline{Coherent superposition and Quantum entanglement}

     It is the possibility of coherent superposition which makes quantum interference effects  possible, possibly the most sensational effects of the so-called wave-particle duality.  
     In Quantum Mechanics, a particle, such as an electron, is modelled by a wave function.   Many physicists have thought that the wave is real, and the wave function represents something real.  There have been many distinguished dissenters, however. For the sake of discussion, let us assume that the wave is  at least equally real as the particle, since  this makes the problem more  difficult to solve, not easier. 

Abstractly, the two-slit experiment boils down to the following, if we put a photodetector directly in one of the two slits.  The state space ${\cal H}_o$ of the particle is spanned by $\psi_1$, meaning it definitely passes through the slit where the detector is, and 
$\psi_o$, meaning it definitely passes through the other slit and fails to be detected.
But any coherent superposition $\psi=c_o\psi_o+c_1\psi_1$ is possible.  The $c_i$ are 
called the quantum amplitudes.  If the detector $M_n$ has n degrees of freedom call its 
Hilbert space ${\cal H}_n$.  If we assume, as Wigner, von Neumann, and practically everybody else do, that the strong topology is physically relevant to the `pointer variable' of $M_n$, then we may assume that one position of the pointer variable, `detection,' the loud click or bright flash, correponds to $|0\rangle\in{\cal H}_n$.  Failure to go off, remaining 
in the charged state of population inversion, is $|1\rangle$.  Then no matter what the 
joint Hamiltonian $H^{com}$ on ${\cal H}_o\otimes{\cal H}_n$ is, it must take 
$\psi_o\otimes|1\rangle$ to $\psi_o\otimes|1\rangle$ and 
$\psi_1\otimes|1\rangle$ to $\psi_1\otimes|0\rangle$.  By linearity, it is forced to take 
$(c_o\psi_o+c_1\psi_1)\otimes|1\rangle$ to $c_o\psi_o\otimes|1\rangle+c_1\psi_1\otimes
|0\rangle$
which is not a decomposable state.  The apparatus and the particle are ``entangled.''
It is no longer acceptable to introduce {\it ad hoc } extra postulates to eliminate this 
entanglement.  

Bell has described the evolution of the coherent superposition 
$c_o\psi_o\otimes|1\rangle+c_1\psi_1\otimes
|0\rangle$
into the statistical mixture (called incoherent)
$$ \vert c_o\vert^2\chi_{\{\psi_o\otimes|1\rangle\}}
+\vert c_1\vert^2\chi_{\{\psi_1\otimes|0\rangle\}}$$
as ``the philosopher's stone'' of quantum measurement.  Here $\chi$ means the characteristic function of a set, and our mixture is a atomic probability density on a two point space.  Hence, the $|c_i\vert^2$ are classical probabilities. 

Thus the problem of Quantum Measurement is as follows:  how do the quantum amplitudes ever get turned into classical  probabilities?  Schroedinger's equation is a linear wave equation, a superposition of two states $\psi_1$ and $\psi_2$ with quantum amplitudes $c_1$ and $c_2$, evolves forever for all future time as a coherent  superposition  with the same quantum amplitudes of two continuously, smoothly evolving quantum states.   Hence, if the incident particle is in such a superposition, so is the combined  system of particle {\it cum } apparatus, and it remains such, forever.  It {\it never }  changes the quantum amplitudes, not even into different quantum amplitudes, much less into probabilities or jumps.  So how can we explain why, when we  analyse the same physical process differently, we obtain that $\vert c_1\vert^2$ and $\vert c_2\vert^2$ are the probabilities that the measurement  apparatus can be observed to be in either one classical pointer position or in  the other?  
 
It would go against the grain of the experimental evidence (that entanglement persists) to 
try to derive this as a time evolution.  We, instead, derive this as a thermodynamic limit.

Quantum amplitudes are essentially quantum mechanical, wave phenomena.  They  are preserved by Schroedinger's equation.  Classical probabilities can be  calculated from the amplitudes, and are observed when we measure, but only  some sort of `cut' between the use of the wave equation and the use of the  probability rules can tell us when to use one and when to use the other.    

\centerline{Where do we put the cut?}

Wigner said, put the cut to separate matter from consciousness.  Bohr and  von Neumann said, put the cut in between the system being measured and the  classically describable, macroscopic measuring apparatus.  Von Neumann went  further and said that if you had another measuring apparatus to measure the  first measuring apparatus, you could go ahead and shift the cut so that the  joint wave function of incident particle {\it cum } first apparatus was being  treated as a wave, smoothly, but any measurement of it by the second apparatus  introduced the jumps.  The praxis of the average physicist in the thirties  just said, put the cut somewhere in the mesoscopic realm so it divides  small molecules from our amplifiers.

The solution which I have proposed is as follows.  We shift the cut out to  infinity.  It only occurs in the sort of idealised limit which is used in  Classical Thermodynamics and probability theory.  No real object is macroscopic, nevertheless, if the number of particles or degrees of freedom in it is quite  large, the idealised limit of macroscopy is a physically valid approximation. No real object behaves exactly according to the axioms governing measuring processes.  But the approximation is, as usual in Classical Thermodynamics,  exceedingly good.  An idealised limit object would behave exactly according to  the axioms, and we can derive its properties mathematically from Schroedinger's  equation simply by passing to the thermodynamic limit in the way usual in  logically careful accounts of classical Statistical Mechanics such as those  by Darwin--Fowler,[4] Khintchine,[12] and Ford--Kac--Mazur.[7,13]  Thus we can derive  the probability rules of Quantum Mechanics from the deterministic rules of  Quantum Mechanics by passing to the thermodynamic limit.              The program, as has existed for a long time now, is to write down the wave equation for the  joint system of incident particle plus amplifying detector.[11]  As one lets  $n$, the number of particles in the detector, go to infinity, the pointer  positions of the detector behave approximately classically, and the approximation  improves the more $n$ increases.  In the limit, the detector is a classical  object, there are no superpositions of pointer positions, and the usual  measurement rules of Quantum Mechanics are exactly true.  But at every point  short of the limit, Schroedinger's equation is instead exactly true, and the  continuous, deterministic evolution of the total wave is also true.  But this  is the same as in Ford--Kac--Mazur's paper (except that they were not thinking about Quantum Measurement, so they were not thinking about amplifying devices, so they only studied positive temperature and not negative temperature), it is a standard feature of classical Statistical Mechanics.  Which is what led Einstein to ask, once, whether the probabilities of Quantum Mechanics did not arise out  of the deterministic dynamics in an analogous way to the way they did in classical Statistical Mechanics (including Thermodynamics as a special case). He seems to have thought that Quantum Mechanics was incomplete [5], and would have to be completed in some radical fashion for this to be true, but Schroedinger  did not think so.  And it turns out that it is not {\it necessary } to alter the  usual wave equation in order to answer Einstein's question in the affirmative. If future experimental phenomena uncover new physical laws which require the  modification of Schroedinger's equation, that would be a different matter.   We need not postulate unmotivated modifications to Schroedinger's  equation just to solve the problem of Quantum Measurement.  Any new equations discovered to explain new physical forces will probably  be susceptible to the same thermodynamic limit treatment, allowing the same  sort of solution to the problem of measurement.

We cannot, naively, pass directly to the limit object by taking the tensor product of 
an infinite number of copies of ${\cal H}_o$.  Although this non-separable space 
does have separable Hilbert subspaces, they are not preserved by the dynamics of the 
measurement process.  Nor is this limit `classical'.  Techniques [2] useful for infinite 
volume quantum limits are not applicable to this physical situation.  
Farhi, Goldstone, and Gutmann [6] have attempted to derive the probabilistic axioms 
from the deterministic axioms, without using statistical mechanical techniques, also leaving aside the reduction of the wave packet.
But there is a hidden assumption in their argument, which tacitly assumes that the results of a measurement 
process obey the laws of probability.  They prove that law's uniqueness, but not existence.  

\centerline{The physical content of the solution}

We take Quantum Mechanics as a complete description of Nature, and  Schroedinger's linear wave-equation as exact, yielding a unitary  deterministic time-evolution for any system not subject to outside  influences.  Since nothing physical is outside Nature, anytime we  wish to improve our approximation of an actual physical system, all  we have to do is include more of the forces acting on it.  This includes  the analysis of measurement processes.  Hence, all systems are taken as closed systems, and only time-independent Hamiltonians are considered.  Furthermore,  all states are taken as pure, i.e., given by wave functions.  

Measuring devices have to be large enough so that they may be approximately  treated as macroscopic, and have to involve amplification and a coupling to  the system they are measuring.  Amplification of a microscopic effect means  that the apparatus has to be in an unstable state of population inversion  (this insight is due to Schroedinger and Green [8])  like a laser or cloud chamber, ready to undergo a sort of chain reaction at the smallest touch, provided it is the right sort of touch to touch off the  chain reaction!  (This insight is due to Daneri, Loinger, and Prosperi.[3])   (Schwinger [14] has also remarked that the negative temperature  amplification of a laser amplifies quantum motion to the classical  level.)    Sometimes such a reaction is called  discontinuous and irreversible, but this is not true at any finite point  short of the thermodynamic limit, which is an idealisation never to be attained.   (In Classical Thermodynamics, too, the discontinuity of phase transitions are only artifacts of passing to the infinite limit.  I would like to thank Goldstone for helping me see the relevance of this.  Even according to Classical Thermodynamics, there isn't really any such thing  as water or steam.  If you looked at the individual molecule, you could not  tell for sure which state it was in.  Neither temperature nor phases really  exist, they are artifacts of the approximation procedure involved in passing  to the thermodynamic limit.  They are still physically valid conceptual constructs.)

This approximation does not take place in the strong topology, and neither does ours.  
Just which quantum-mechanical property or observable of the incident particle  is measured depends explicitly on the physics of the coupling.
We thus {\it derive } non-commutative probability from ordinary, commutative, Kolmogoroffian probability.  That is, for a fixed experimental set up, the abelian probabilities of ordianry classical statistical mechanics are implemented as an algebra generated by one two-by-two matrix-----thus abelian.  But choice of a second, alternative, incompatible experimental set-up yields a different abelian algebra which would not have commuted with the first alternative.  The two non-commuting quantum probability `variables' can never be both realised at the same time.  Hence quantum probability is of strictly {\it derivative } status from commutative probability and the physics of the coupling.  

Only some features of the measuring apparatus are what we define as macroscopically observable, quantum states which differ by negligible numbers  of occupation numbers (by negligible, we mean in some vanishing proportion in  the thermodynamic limit) are not distinguishable, at least in the limit, by  macroscopic observables.  (This is the exact opposite of Hepp's notion [10] of  macroscopic observable \dots)  Some macroscopic observables are not very  relevant to the analysis of measurement, those that are we call macroscopic  pointer variables.  Very few such exist, and for a given coupling, only  inessential variations on a canonical representative seem to exist.  In the  thermodynamic limit, these pointer variables pass into classical probability  measures on a classical object, the limit of the quantum combined 
system.  

There are only waves.  There are no particles.   The fact that measurements seem to detect particles is an artifact of the process of amplification.  The seeming `particle-events' that occur when we measure a field and detect a quantum are the product of the interaction of the wave functions of the measuring apparatus and the microscopic system being detected.

Since thermodynamic limits do not actually exist in Nature (no amplifying device with an actual infinite number of degrees of freedom {\it relevantly coupled } to the incident particle have yet been observed in Nature), none of the jumps  physically occur, and the probabilities we calculate are merely convenient ways  to talk about the results of experiments, approximately valid because of the  usefulness of thermodynamic limits in calculating good approximations to  expectation values or time averages.  

\centerline{Mathematical content}

The only new feature is the rigourous definition of macroscopic observable.  In 
keeping with our program, these are classical dynamical variables, i.e., abelian 
functions on the phase space.  Let ${\cal H}_n={\cal H}_{n-1}\otimes {\cal H}_0$ be a sequence of state spaces 
with an increasing number of degrees of freedom. Let $f_n$ be a 
real-valued measurable function on ${\cal H}_n$.  Let $\langle f_n\rangle_t$
be the time average of $f_n$ over a trajectory and we assume that this is almost 
independent of the initial condition chosen in the space of accessible states.
The sequence $\{f_n\}$ is a macroscopic observable if it is, so to speak, a tail event.
More precisely, we require that if $\{v_n\}$ is a sequence of elements of ${\cal H}_o$
(of norm one) such that  
$\lim_{n\rightarrow\infty}\langle f_n(v_o\otimes v_1\otimes\dots\otimes v_n)
\rangle_t$
exists, then this limit is independent of finite substitutions in the $v_i$.
This implements the intuition that naked eye observation cannot detect microscopic 
variations in the state and can only detect properties which pass to the thermodynamic 
limit in which Planck's constant is negligible.
Details will appear elsewhere.  

Pointer positions are examples of a macroscopic observable.

\centerline{Bibliography}

[1] J.~S. Bell, On Wave Packet Reduction in the Coleman--Hepp Model, {\it Helv. Phys. Acta} {\bf 48}, 447 (1975); {\it Phys. World} {\bf 3}, 33 (1980).

[2] F. Benatti, {\it Deterministic Chaos in Infinite Quantum Systems}, Trieste 1993.

[3] A. Daneri, D. Loinger and G. Prosperi, Quantum Theory of Measurement and Ergodicity Conditions, {\it Nucl. Phys.} {\bf 33}, 297 (1962).

[4] C.~G. Darwin and R.~H. Fowler, On the Partition of Energy, {\it Phil. Mag.} {\bf 44}, 450, 823 (1922). 

[5] P.~A.~M. Dirac, {\it Directions in Physics}, N.~Y., 1978.  

[6] E. Farhi, J. Goldstone and S. Gutmann, {\it Ann. Phys.} {\bf 192}, 368 (1989).

[7] G.~W. Ford, M. Kac and P. Mazur, Statistical Mechanics of Assemblies of Coupled Oscillators, {\it J. Math. Phys.} {\bf 6}, 504 (1965).

[8] H.~S. Green, Observation in Quantum Mechanics, {\it Nuovo Cimento} {\bf 9}, 880 (1958).

[9] K. Hannabuss, Dilations of a Quantum Measurement, {\it Helv. Phys. Acta} {\bf 57}, 610 (1984); {\it Ann. Phys.} {\bf 239}, 296 (1995).

[10] K. Hepp, Quantum Theory of Measurement and Macroscopic Observables, {\it Helv. Phys. Acta} {\bf 45}, 237 (1972).

[11] J.~F. Johnson, {\it Proc. VIIIth Inter. Wigner Symposium, N. Y. 2003,} edited by Catto and Nicolescu.

[12] A. Khintchine, Mathematical Foundations of Statistical Mechanics, N.~Y. 1949.

[13] J. von Plato, Ergodic Theory, in Skyrms, Harper, ed.s, {\it Causation, Chance, and Credence, Proceedings of the Irvine Conference on Probability and Causation, 1985 vol.\ 1}, edited by Harper and Skyrms, 257, Dordrecht, 1988.

[14] J. Schwinger, Brownian Motion of a Quantum Oscillator, {\it J. Math. Phys.} {\bf 2}, 407 (1961).

[15] E. Wigner, {\it Am. J. Phys.} {\bf 31}, 6 (1963).

[16] W. Zurek, {\it Phys. Rev.} {\bf D24}, 1516 (1981), {\bf D26}, 1862 (1982) ; {\it Physica Scripta} {\bf 76}, 186 (1998).

\end